**Research**

# A price on warming with a supply chain directed market

John F. Raffensperger[1]

**Abstract**
Existing emissions trading system (ETS) designs inhibit emissions but do not constrain warming to any fixed level, preventing certainty of the global path of warming. Instead, they have the indirect objective of reducing emissions. They provide poor future price information. And they have high transaction costs for implementation, requiring treaties and laws. To address these shortcomings, this paper proposes a novel double-sided auction mechanism of emissions permits and sequestration contracts tied to temperature. This mechanism constrains warming for many (e.g., 150) years into the future and every auction would provide price information for this time range. In addition, this paper proposes a set of market rules and a bottom-up implementation path. A coalition of businesses begin implementation with jurisdictions joining as they are ready. The combination of the selected market rules and the proposed implementation path appear to incentivize participation. This design appears to be closer to "first best" with a lower cost of mitigation than any in the literature, while increasing the certainty of avoiding catastrophic warming. This design should also have a faster pathway to implementation. A numerical simulation shows surprising results, e.g., that static prices are wrong, prices should evolve over time in a way that contradicts other recent proposals, and "global warming potential" as used in existing ETSs are generally erroneous.

**Keywords**  Emissions trading systems · Global warming

## 1 The flaw in existing market designs for emissions

Despite the urgent need to reduce greenhouse gas (GHG) emissions, we all have no certainty of any particular pathway of emissions. We lack pathway certainty because emissions have no contractual adherence to any fixed global limits. Worse still, we have no certainty of any pathway of warming. To address these problems, this paper proposes a centralized global market for permits to emit GHGs and for contracts to sequester carbon dioxide ($CO_2$). The proposed market explicitly constrains warming in degrees Celsius by time period far into the future. The market sets prices which result in a specified warming pathway. Further, this paper proposes a set of market rules and a bottom-up implementation path of willing firms working through their supply chains and jurisdictions joining as they are ready. This market appears to have bottom-up incentives for joining and these incentives increase over time under realistic conditions.

The scientific community knows the flaws in the various emissons trading systems (ETSs). Sadly, jurisdictions have applied formal pricing mechanisms to less than 20% of the world's emissions [1]. Where emissions markets exist, they do not always guarantee any particular limit (e.g., Australia's "Safeguard" mechanism). In a well-regarded proposal, Wilcoxen and McKibbin [2] proposed caps by country with international trading: "Because the total supply of permits in any year would not be fixed, the policy would not guarantee precisely how much abatement will take place in that year." Where ETSs do attempt to reach a particular limit, the limit can be subverted (e.g., California; [3]),

---

and the limit does not apply to certain sectors (such as agriculture in New Zealand). ETSs reveal allowance prices, but policymakers and researchers can only estimate impacts on emissions [4]. Emissions markets can end on political whims (e.g., Ontario and Australia). Credit banking allows companies to buy the right to emit now, but actually emit later, lowering pathway certainty. Inconsistent metrics ("global warming potential" in particular) ignore the different warming effects for emission of a given chemical over time [5, 6]. Some market rules incentivize cheating; for example, paying to destroy stocks of chemicals incentivizes covert production of those chemicals. See World Bank [7] for an up-to-date report on carbon pricing, and Stavins [8] for a careful analysis of the full landscape of pricing policy.

Ideally, the international community would come together to set up a global emissions trading scheme (ETS). Ideally, jurisdictions would agree to firm uniform rules, selection of the warming pathway, how to improve the science over time, enforcement of the rules, and supervision of the market manager. Yet, the world has not yet found a path toward this ideal institution due to the transaction costs and weak incentives for setting it up.

Compared to cap-and-trade markets, policymakers and researchers believe carbon taxes are easier politically, especially when combined with "dividends" to consumers. See, in particular, Shultz and Halstead [9] who focus on political expediency, and Weisbach and Metcalf [10] who thoughtfully and optimistically discuss the difficulties of tax design in good detail. Kaufman et al. [11] propose that policymakers figure out targets to "net zero" $CO_2$ emissions; their proposal seems no easier politically than other cap and trade mechanisms. Carbon taxes (as in Argentina) do not reduce emissions to any specific limit. Even a uniform global tax or a uniform price floor (e.g., [12]) could not guarantee the correct pathway of emissions, much less warming, because no policymaker can find the correct price for enforcing the required pathways of either emissions or warming. If the tax undercharges, we make insufficient progress. If the tax overcharges, we inhibit growth needlessly [13]. For emitting firms, transaction costs with carbon taxes are lower than with cap and trade [10]. But vested interests block implementation of carbon taxes and overcoming those interests is an insurmountable transaction cost.

Further, carbon taxes have no inherent mechanism for the vast sequestration required to remove the excess $CO_2$. Rather than establishing a robust procurement mechanism for sequestration, the money would go to bribing voters (increasing consumption), or subsidizing technologies (likely misguided, as "carbon capture and sequestration"), or building infrastructure (increasing consumption), or some other government priority. Carbon taxes would more effectively reduce warming if policymakers used the money to procure sequestration.

Thus, no existing emissions market design constrains warming to any particular managed path when constraining warming is the whole point. The result is *management* uncertainty about emissions on top of and far greater than the scientific uncertainty of the warming those emissions will bring. We all know GHG emissions cause global warming. The biggest uncertainty is not the warming of a ton of any gas but rather what we will do about it. I call this management uncertainty because the human race as managers of our emissions can control it. A system that controls emissions with more management uncertainty has more uncertainty than a system that controls emissions with less management uncertainty. To quote Stiglitz [14]:

> In the context of climate change, there is considerable uncertainty, e.g., about the magnitude of the links between greenhouse gases and climate change and that of the links between any instrument and greenhouse gas emissions. The latter uncertainty has led some environmentalists to argue for quantitative restrictions on emissions… One way of understanding this is to note that while the standard result argues for a single price of carbon in all places, for all uses, at all dates, the (appropriate shadow) price differ depending on the state of the world. There is much we don't know: the effects of any policy on emissions or the effects of emissions and carbon concentrations on climate change, and the full effects of climate change on well-being. Thus, as we learn more about the state of the world the carbon price adjusts. In fact, the best we can do is to announce a carbon price today and a limited state-dependent sequence of prices going forward.

The proposal here aims to design an auction mechanism which at every clearing announces both a current price and a temperature-dependent sequence of future prices far into the future. The mechanism creates market forces to help agents anticipate changes in learning and technology, which policymakers cannot foresee. In addition, the proposal here is for a group of firms to begin a credible ETS as a type of climate club [15, 16]. The difference from previous proposals is the path to get there, which does not require governments to make agreements. Market rules specify sanctions against non-participating firms from both member jurisdictions and participating firms, including those in non-member jurisdictions. Whether a country joins or not, it avoids a politically risky commitment to emission targets. Centralization allows firms to participate though the jurisdiction has not formally joined. This feature could result in de facto membership of a jurisdiction, lowering the political barrier for the jurisdiction to join. The result is bottom-up incentives for joining, and these incentives increase over time under plausible conditions, potentially resulting in a broad rush to join while inhibiting dropouts.

Section 2 describes the centralized auction. Clearing the auction with optimization enables the auction manager to find prices as a function of warming effects, for current and future periods, at every auction. Section 3 describes the market institution's basic rules, incentives, governance and enforcement. Section 4 presents results of a numerical simulation of the mechanism.

Section 5 proposes a faster path forward in which the coalition of businesses drives the establishment of this new ETS through their supply chains, with lower transaction costs of building the institution compared to achieving international agreements. This section also introduces the "CLEP"—customer-labeled emissions permit which a firm can use to prove to customers that the firm has purchased emissions credits for their products. Section 6 is the conclusion.

## 2 A mechanism for a price on warming

### 2.1 The smart market approach

A smart market is a centralized auction cleared by optimization. The optimization enables trading of contracts in which the traded commodities have complex physics and shared constraints which otherwise hinder trading. A smart market enables the type of central control instrument which Weitzman [17] called "ideal" but also "infeasible," because transmitting "an entire schedule of ideal prices or quantities" contingent on the state of the world requires "a complicated, specialized contract which is expensive to draw up and hard to understand." A smart market is not a price mechanism like a tax, nor a quantity mechanism like cap-and-trade, but rather is simultaneously a price and quantity mechanism. With roots in operations research going back to Dantzig in the 1940s, the smart market concept was developed in the late 1970s and 1980s [18].

Smart markets are in active use for wholesale electricity and natural gas, transportation, and radio spectrum. Similar mechanisms are in use for medical internships, organ transplants, and specialized procurement. Those applications require allocations to adhere to complicated shared constraints which otherwise impose transaction costs in a normal market. The smart market drastically reduces these transaction costs by incorporating the shared constraints in an optimization used to clear the market. In the case of wholesale electricity, for example, generation must match demand within various system capacities, which the electricity system operator incorporates into a linear program to clear the market. In the proposed design here, the shared constraints limit warming for each time period far into the future. To be clear, this is not a one-shot auction, but rather recurring every period, e.g., every month.

I call this market design and the associated optimization model SMDAMAGE, for Smart Market Design Addressing Management of Global Emissions. The market would sell contracts for permits to produce GHG chemicals and buy contracts for $CO_2$-sequestering activities. Participants trade through a central market manager: buyers buy from the market manager and sellers sell to the market manager. The market manager is responsible for registering agents to trade, for enforcing market rules, and for proper functioning of market operations. Agents could sell back unused permits and buy back inactive sequestration contracts.

Since this mechanism does not impose quantities in a regulatory way and since the price mechanism is not a fixed tax, the smart market seems to be what economists are calling a hybrid system, though it is simpler than those described in the literature [19].

This mechanism does not attempt to incorporate the social cost of carbon. Kaufman et al. [11] concluded that the social cost of carbon is too far removed from actual policy-making. In SMDAMAGE, the auction incorporates only bids from buyers of emission permits and sellers of sequestration contracts. The underlying externality is warming itself, which the auction manages explicitly.

### 2.2 Nature of the traded contracts

The proposed market trades two types of contracts. The first type specifies the right to produce a quantity of GHG chemical during a certain time. The second type specifies the obligation to sequester carbon dioxide during a certain time. Trading contracts of multiple types in an ETS is common [20].

A contract to produce specifies a chemical, a quantity (e.g., in kilotons), and specific start and end times, e.g., 1 month. For simplicity of explanation, assume the agent produces the chemical and immediately emits it. In reality, the contract would specify the timing of emission. Payment for production moves the transaction cost upstream from the consumer, increasing economies of scale by lowering administrative costs [21].

Buyers obtain the permission, but not the obligation, to produce the quantity of pollutant for the scheduled interval of the contract. Agents can buy permits for future periods, as far as the final bid period $T_B$ in the market schedule. The term length of the typical contract for production would be short, e.g., a month. Buyers cannot apply the contract to a different schedule than the one listed. They can use the contract to produce only for the schedule listed on the contract. If the buyer does not produce during the specified time, the permit expires with no value.

Agents might need to buy more than one kind of right at a time. For example, consider an agent operating a natural gas combined cycle plant for power generation with carbon capture and storage. The agent should buy permits for uncaptured carbon dioxide production and also permits for methane leakage.

Agents who buy future production permits may offer to sell them back to the market. For example, consider a generator holding methane permits for 5 years into the future. The generator could reduce methane leakage early and then sell the remaining methane permits back to the market manager. The generator has no guarantee the selling price would be the same as the original buying price. The generator could also sell the contract privately, but would have to register the trade with the market manager; sale through the central auction should have a lower transaction cost.

A contract to sequester specifies an activity, start and end times, and kilograms quantity in each period following a specified schedule. These contracts might be complicated. Sellers accept the obligation to sequester from the atmosphere the quantity of carbon dioxide over the agreed time. The term length of the typical contract for sequestration can vary depending on the activity, and agents can sell sequestration services for future periods, as far as the final bid period $T_B$ in the market schedule. For example, a contract to grow a hectare of pine trees might have a term of 30 years, absorbing different amounts of carbon dioxide each year. Table 1 shows example contracts for a fictional 2021 auction.

Agents who sell future sequestration contracts to the market manager might offer to buy them back from the market manager. For example, suppose an agent has sold a 30-year contract to grow a hectare of forest, but 15 years into the contract, wants to change the forest to some other crop, which sequesters less carbon dioxide. This agent could offer to buy the remaining 15 years of the contract back from the market manager.

The final period of constraint must be far in the future, e.g., 150 years, depending on the residence time of the different chemicals in the atmosphere. For various reasons, especially moral hazard and adverse selection, contracts for sequestration probably should not extend more than a few decades. The 150-year period corresponds to constraints on warming from near-term contracts. To account for these varying warming trajectories, the market manager chooses bids with an optimization model.

### 2.3 Market clearing model SMDAMAGE

SMDAMAGE is a linear program. Computers can solve it easily and the solution algorithm produces price information [22]. For now, assume the warming effects of different chemicals combine in linear and deterministic fashion. The model allows the possibility a chemical could change to a different compound, with a different effect on global warming. Other researchers might be able to find a model with better accounting for chemical interactions and feedback loops. This research should be ongoing, but implementation should not wait for more research. Instead, market rules should allow for improving the science.

#### 2.3.1 Indices

$a$ = agent, $a = 1,\ldots, A$.

**Table 1** Example contracts traded in a fictional SMDAMAGE smart market held in 2021

| Type | Term | Quantity | Price |
|---|---|---|---|
| $CO_2$ emission | 2040-06-01 to 08-31 | + 100 ktons $CO_2$ | $32.50/ton |
| $SF_6$ emission | 2024-01-01 to 03-31 | + 0.5 ktons $SF_6$ | $51,200/ton |
| Agriculture, biochar | 2025-04-01 to 2030-03-31 | − 15 ktons $CO_2$ | − $32.50/ton |
| Forestry, Loblolly pine | 2023-06-01 to 2053-05-31 | Varying amounts over 30 years | − $3340/hectare |

$p$ = activity (pollutant or contract), $p = 1,\ldots, P$. For sequestration contracts, $p$ is $CO_2$, though the market manager may specify the contract in other units, such as hectares of forest.

$t, u$ = period, where $t, u = 1,\ldots, T$, where $T \approx 7500$ weeks. Generally, subscript $u$ indicates the period of emission and $t$ to indicate the period of warming.

### 2.3.2 Parameters

$Cap_t$ = allowed increase in thousandths of degrees Celsius in period $t$, relative to some baseline temperature. Critical values for the auction include the start date of the first auction and the first period $Y$ of constrained warming. The trajectory of this crucial parameter could follow scientific recommendations, but the managing committee might wish to change $Cap_t$ as future impacts become clear. $Cap_t > 0$ allows a global temperature above the baseline in period $t$. $Cap_t = 0$ requires the global temperature in period $t$ to equal the global baseline temperature. $Cap_t < 0$ requires a fall in the global temperature in period $t$, e.g., to the global temperature of 1920, in which case the market manager likely has a net buying position to get sufficient sequestration to reduce the global temperature.

$B_{a,p,u}$ = bid price by agent $a$ to produce 1 unit of activity $p$ in period $u$, e.g., a kiloton of $SF_6$ released or a hectare of forest to sequester $CO_2$. The market manager could mark up this bid price by the cost to enforce the contract. Agents are responsible for discounting their own bid prices.

$Q_{a,p,u}$ = upper bid quantity by agent $a$, units of activity $p$ (e.g., kilotons $CO_2$ or hectares forest) in period $u$.

$T$ = final year that $Cap_t$ is constrained, e.g., the year 2301. This is the end of the auction horizon specified by the market manager.

$T_B$ = final year of denomination for contracts in the current year, specified by the market manager. For example, an auction held in 2021 would allow trading of contracts through $T_B = 2050$, but not thereafter. To partially address adverse selection, the auction manager could buy sequestration contracts for only the near term (implying one $T_B$ for emissions and a shorter $T_B$ for sequestration). I address this more later.

$Y$ = first year in which $Cap_t$ is constrained: $Cap_t = \infty$ for $t < Y$ and $Cap_t = 0$ for $t \geq Y$. The market manager specifies this parameter. For example, an auction held in 2021 could have no constraints for temperature up to $Y = 2040$, and then constrain temperature from $Y = 2040$ through $T = 2301$. Because of storage in the atmosphere, prices in 2021 will immediately drive both emission reduction and sequestration.

$I_p$ = the initial atmospheric burden of pollutant $p$, in units (e.g., megatons or kilotons) consistent with $W_{p,u,t}$ above a base value, such as the difference between 2021 and 1920 quantities.

$W_{p,u,t}$ = absolute global temperature change in period $t$ for one kilogram of pollutant $p$ released or $CO_2$ sequestered from activity $p$ in period $u$, where $u \leq t$ [5, 23]. $W_{p,u,t}$ is negative for a sequestering activity.

Figure 1 shows $W_{p,u,t}$ schematically for a generic pollutant. $W_{p,u,t}$ might be positive in early years and negative in later years, or vice versa, depending on activity $p$. The factor can account for chemical changes, such as methane changing to $CO_2$. These effects are somewhat uncertain, so this value should be set conservatively, e.g., at the 99th percentile of certainty. We can define more complicated contracts. For example, for a hectare of pine trees planted in period $u$, denote the schedule of tons $CO_2$ sequestered as $N_u, N_{u+1}, N_{u+2}, N_{u+3}, \ldots, N_T$. Then a period $u$ contract for one hectare will warm period $t$ by $W_{pine,u,t} = -\sum_{v=u}^{t} N_v W_{CO2,v,t}$. This formula can account for specific types of trees and for seasonal effects. Thus, the formulation is general.

### 2.3.3 Decision variables

$q_{a,p,t}$ = kilograms allocated to agent $a$ to produce activity (e.g., pollutant or sequestration contract) $p$ in period $t$.

$v_{p,t}$ = total activity $p$ in period $t$. We can interpret this variable as the total emissions or sequestration by sector $p$.

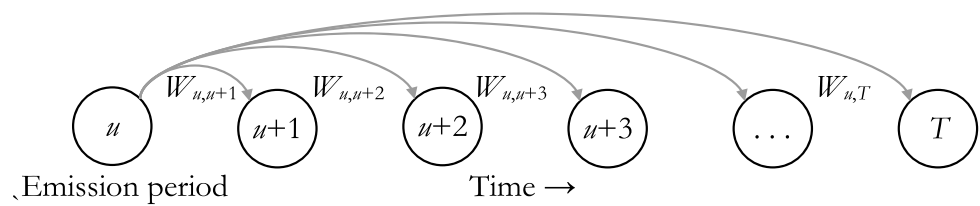

**Fig. 1** Schematic of $W_{u,t}$, the future warming effects of a generic pollutant emitted in period $u$

$\pi_{p,t}$ = market price, $ per unit (e.g., kilotons or hectares) of activity $p$ in period $t$. This is the dual price (also called shadow price, marginal price, Lagrangean value) on constraint 3 below.

$\omega_t$ = marginal cost for a change in $Cap_t$. This is the dual price on constraint 4 below. We can interpret $\omega_t$ as the improvement to the market manager's net revenue for an increase in the cap on warming in period $t$.

### 2.3.4 Model SMDAMAGE

$$\text{Maximize} \sum_{\text{agent } a=1}^{A} \sum_{\text{activity } p=1}^{P} \sum_{\text{period } t=1}^{T_B} B_{a,p,t} q_{a,p,t}, \tag{1}$$

$$q_{a,p,t} \leq Q_{a,p,t}, \text{ for agent } a = 1, \ldots, A, \text{ activity } p = 1, \ldots, P, \text{ and period } t = 1, \ldots, T_B, \tag{2}$$

$$\sum_{\text{agent } a=1}^{A} q_{a,p,t} = v_{p,t}, \text{ for activity } p = 1, \ldots, P, \text{ and period } t = 1, \ldots, T_B, \text{ dual price } \pi_{p,t}, \tag{3}$$

$$\sum_{\text{activity } p=1}^{P} W_{p,0,t} I_p + \sum_{\text{activity } p=1}^{P} \sum_{\text{emission periods } u=0}^{t} W_{p,u,t} v_{p,u} \leq Cap_t, \text{ for } t = Y, Y+1, Y+2, \ldots, T, \text{ dual price } \omega_t, \tag{4}$$

$$q_{a,p,t} \geq 0 \text{ for all agents } a, \text{ activities } p, \text{ and periods } t. \tag{5}$$

### 2.3.5 Explanation

1. Maximize the value of the traded contracts to market participants. If the market manager based prices on bids rather than the marginal, the objective would maximize the market manager's profit. This accepts the highest bidders for permits and the lowest offers for sequestration, subject to the constraints.
2. Respect agents' upper bid limits, as specified in their bids.
3. Calculate the total quantity of activity $p$ each period as the sum of the agents' allocated quantities. The dual price $\pi_{p,t}$ on constraint 3 serves as the global price for activity $p$ in period $t$. This price accounts for participants' bids for the activity, the residence time of the pollutant activity, and the warming cap in each period.
4. The market model caps the cumulative warming effects of emissions for period $Y$ and thereafter in the auction calendar. Therefore, the first auction will calculate prices for every future period in the auction calendar. Those prices could change at the next auction. SMDAMAGE allows flexibility with this constraint, e.g., to constrain the number of degree-months above $Cap_t$.
5. Variables $q_{a,p,t}$ are nonnegative. The model does not need non-negativity constraints for $v_{p,t}$ because Eqs. 3 and 5 together ensure the nonnegativity of $v_{p,t}$.

### 2.3.6 Market clearing

Before bidding, agents wishing to trade must register with the market manager, verifying a range of fiduciary attributes. The market manager has the right to reject an agent for misbehavior, lack of credit worthiness, etc.

At the beginning of a market cycle, the market manager would open a bidding page like Fig. 2. While the bidding page is open, perhaps up to a few hours before each auction, registered agents would enter their bids as price and quantity pairs. Figure 2 shows a sketch of the bidding webpage, as a fictional bidder $a$ = Emisor DeGas, Ltd., might fill it out. First the bidder selects the activity, in this case $p$ = nitrogen dioxide. Second, the bidder selects the start date of emission $u$ = 2021-08-01. Third, the bidder selects the bid price $B_{a,p,t}$ and the maximum quantity $Q_{a,p,t}$. Finally, the bidder clicks the submit button.

The price for a buy bid is the highest price the agent would be willing to pay for the given quantity. A buyer might get none, some, or all of the bid quantity. The price for a sell bid is the lowest price the agent would be willing to sell the given quantity. The market manager might accept none, some, or all of the seller's quantity. To protect proprietary data,

**Bid for Emisor DeGas, Ltd.**

| a. Choose permit type | b. Choose emission period August 2021 | | | | | c. Enter bid price and quantity. | |
|---|---|---|---|---|---|---|---|
| Carbon dioxide, ktons | | | | | | Nitrogen oxide, ktons | Action $p$ = NO2 |
| Methane, ktons | Sun | Mon | Tue | Wed | ... | 2021-08-01 | Period $u$ = 2021-08-01 |
| Sulfur hexafluoride, ktons | 31 | 1 | 2 | 3 | ... | $1,115.00 | B[DeGas, NO2, 2021-08-01] |
| Nitrogen oxide, ktons | 7 | 8 | 9 | 10 | ... | 80 ktons | Q[DeGas, NO2, 2021-08-01] |
| Pine trees, hectares | 14 | 15 | 16 | 17 | ... | Submit bid | |

**Fig. 2** Sketch of the bidding web page, showing parameters at right

bids are private to the agents and the market manager. The market manager does not need to guess a discount rate, because participants use their own.

At the preannounced time, the market manager would close the bidding page. The market manager would construct the linear program SMDAMAGE and solve it. The market manager accepts bids specified by the decision variables $q_{a,p,t}$ and announces the prices $\pi_{p,t}$ for each activity $p$ and each period $t$. Thus, the market follows marginal cost pricing. All prices $\pi_{p,t}$ are public. The market manager authorizes payment from buyers and authorizes payment to sellers (Fig. 3).

The following formula gives the market manager's revenue:

$$\text{Revenue in the current auction} = \sum_{\text{agent } a=1}^{A} \sum_{\text{activity } p=1}^{P} \sum_{\text{period } t=1}^{T_B} q_{a,p,t} \pi_{p,t} \text{sgn}(B_{a,p,t}) \qquad (6)$$

where $\text{sgn}(B_{a,p,t}) = -1$ if $B_{a,p,t} < 0$ when agent $a$ sells a sequestration contract, else $\text{sgn}(B_{a,p,t}) = 1$ when agent $a$ buys emissions permits.

After the first auction, global markets would observe posted prices for each warming activity for each period for decades into the future. This price information would provide important signals to business and governments. The warming effects $W_{p,u,t}$ of each activity $p$ into the future drive the market prices $\pi_{p,u}$ as seen from the mathematical dual constraint associated with the primal decision variable $v_{p,u}$.

$$\sum_{\text{periods } t=u}^{T} W_{p,u,t} \omega_t - \pi_{p,u} \geq 0, \text{ activity } p = 1, \ldots, P, \text{ and emission period } u = 1, \ldots, T_B, \text{ primal variable } v_{p,u} \qquad (7)$$

Thus, the market price in any period for emitting each activity depends on the activity's future warming effects and the desired management pathway for warming. Such advance prices would better incentivize long-term planning of technologies [24].

## 3 Basic rules and incentives of the market

### 3.1 Basic rules: universal participation and multilevel enforcement

In this proposal, all jurisdictions (countries, states, provinces, etc.) require participation of all emitters of GHGs within their jurisdictions. All production anywhere of the listed GHGs requires purchase of a contract through this market. Jurisdictions would consolidate other emissions markets into this one. Market rules would prohibit grandfathering.

Responsibilities for enforcement would be multilevel:

**Fig. 3** Sketch of bidding web page results for fictional agent Emisor DeGas, Ltd

> **Bid Conclusion for Emisor DeGas, Ltd.**
> The market has cleared. Your bid was accepted for:
> q[DeGas, NO2, 2021-08-01] = 80 ktons NO2
> at π[NO2, 2021-08-01] = $1,048.54/kton
> for emission period starting 2021-08-01.

| | |
|---|---|
| Rule 1 | The governing committee chooses the global limits $Cap_t$ on warming. |
| Rule 2 | Participating commercial agents have responsibility to obtain permits for all production. |
| Rule 3 | Everyone has responsibilities to avoid trade with non-participating agents, even if the participating agent's jurisdiction is a non-member. |
| Rule 4 | The governing committee has responsibility for promoting individual concern worldwide for everyone to support participating companies and to boycott companies that do not. |
| Rule 5 | The market manager has responsibility to maintain a public list of non-member jurisdictions, and a public list of commercial agents who do not participate or who violate the rules. |
| Rule 6 | The market manager has responsibility for paying whistle-blowers who expose violations. |
| Rule 7 | Each member jurisdiction has responsibility for requiring agents within its jurisdiction to produce no more than permitted. |
| Rule 8 | Member jurisdictions have the right to serve on the governing committee. Member jurisdictions found to have lax enforcement could lose committee voting rights. |
| Rule 9 | Each member jurisdiction has responsibility to impose trade sanctions against non-member jurisdictions, following Nordhaus [15]. |
| Rule 10 | Each member jurisdiction has responsibility to sanction trade specifically against non-participating agents wherever they produce. This includes preventing or taxing trade with non-participating agents, and waiving tariffs on participating agents who produce in nonmember jurisdictions. |

Note the difference in rules to other market designs: participating commercial agents have responsibility to avoid trade with nonparticipating agents without regard to jurisdiction. To my knowledge, no other ETS has such a rule.

The insight here is that company-level supply chain incentives are a critical part of the solution. Companies' incentives are heterogeneous. Taking advantage of heterogeneity can create a path forward and strengthen enforcement. Some businesses badly want to offset their emissions and these firms can lead the way using the mechanism described here [25].

### 3.2 Governance and enforcement

The market should cover at least the GHGs listed in the Kyoto Protocol, including carbon dioxide, hydrofluorocarbons, methane, nitrogen trifluoride, nitrous oxide, perfluorocarbons, and sulfur hexafluoride. Market rules should allow the governing committee to add other GHG chemicals to the auction.

Governance should include improvement of the operation of the market, such as reducing transaction costs ([26], but see [27], for a different view), especially for sequestration projects [28, 29].

Because the first principle of this proposal is adherence to the warming pathway, the market manager and jurisdictions could use the first tranche of revenue for enforcement. The market manager could use revenue from fines to retire production permits. If the manager had a policy of retiring more permits than necessary to offset the emissions from the infraction, the threat of lower supply might incentivize other emitters to report infractions.

The market manager should not be the sole enforcer. Jurisdictions must also be responsible for local enforcement. With a strong whistle-blower policy, whistle-blowers could be entitled to payments based on current market prices. The governing committee should promote a global attitude of individual support and personal responsibility for enforcement.

With the market rules allowing for improving science over time, regulators should also improve enforcement over time. For example, jurisdictions and the market manager may demand blockchain corroboration of agents' emissions sensors. Poor jurisdictions might need help with enforcement.

Market rules would prohibit the market manager from paying anyone to avoid emissions. The market manager can pay participants only for sequestration. The governing committee will have to establish rules for existing stocks.

Market rules would establish a process for improving the science over time, as well as how to adjust previously allocated permits when that science results in changes to parameters $W_{p,u,t}$ or $Cap_t$. Other researchers might find an alternative formulation of SMDAMAGE with better modeling of the physics. The basic design of the smart market would remain the same.

### 3.3 Incentives to join for jurisdictions and to participate for commercial agents

The proposed centralized auction with decentralized enforcement appears to incentivize jurisdictions to join and commercial agents to participate. The key difference to other designs is the ability of firms to participate even if their jurisdictions are non-members, combined with members and participants fostering broad societal pressure on non-member jurisdictions and non-participating firms. Centralization enables this feature, unlike regional ETSs. Further, these incentives appear to increase over time. These characteristics could result in a cascading bottom-up "me too" rush to join, in sharp contrast to the "you first" incentives of other designs.

Determining the true willingness of firms to participate would take considerable effort. Their true willingness is probably unknowable until they actually put up money; a survey may simply find cheap talk. We can only speculate based on published reports (e.g., Bloomberg [30]) that some firms do want to participate. So assume a coalition of willing firms and jurisdictions starts the market. Some firms might agree to participate even if their own jurisdictions do not join. Remaining jurisdictions are non-members and their commercial agents are non-participants. The market manager auctions all permits, holding nothing back.

Consider the incentives of a non-participating commercial agent within a non-member jurisdiction, such as the firm represented by a circle at bottom left in Fig. 4. The market manager would record the agent as a scofflaw. First, the agent faces increasing trade sanctions by member jurisdictions and participating agents, especially if the agent has significant export operations. The agent may participate to avoid sanctions. Second, the agent might expect their jurisdiction will eventually become a member, at which point the agent would need production permits. The agent has incentive to participate even before its jurisdiction joins to be ready for this contingency and to avoid possible price increases. Because the market manager puts all permits on sale, this availability might incentivize participants to join early in hopes of lower prices, which are likely to rise as more participants join. Third, the non-participating agent faces reputational risk with its customers, local and international stakeholders, and whistle-blowers. This risk would increase over time as news of the agent's non-participation became more widely known. Thus, as more agents participate, a given non-participant in a non-member jurisdiction faces increasing business risk, increasing trade sanctions, increasing reputational risk, and increasing likelihood their own jurisdiction will join.

Consider the incentives of a non-member jurisdiction, such as the jurisdiction represented by the box at the bottom right of Fig. 4. First, participation of commercial agents within could result in de facto jurisdictional membership, though the jurisdiction is a non-member. Joining would become politically easier over time. Second, the non-member jurisdiction would have no right to auction management though an increasing revenue comes from agents within its borders. A variety of stakeholders within the non-member jurisdiction would then advocate for the jurisdiction to join. Third, the jurisdiction could save money by eliminating subsidies and regulations made obsolete by the central market. Thus, as more agents anywhere participate, the non-member jurisdiction faces a decreasing political risk to join and an increasing political risk to stay out. When the non-member decides to join, all commercial agents within its jurisdiction would be required to participate, raising the incentives even further for other agents to participate and for other non-member jurisdictions to join.

Consider the incentives of commercial agents who cheat. Producers of GHGs with the highest permit prices, such as sulfur hexafluoride or nitrous oxide, will have correspondingly high incentive to cheat. Cheaters might produce more than permitted or produce at different times than permitted. Cheaters might trade with non-participating agents. The market manager would be primarily responsible for publicizing the identities of cheating agents. However, all member jurisdictions and participating agents would be responsible for detecting cheaters and imposing sanctions on them. Whistleblowers could expose the scofflaw. As an increasing number of agents participate and jurisdictions join, sanctions against cheaters would increase. The incentive for commercial agents to cheat appears to decrease over time.

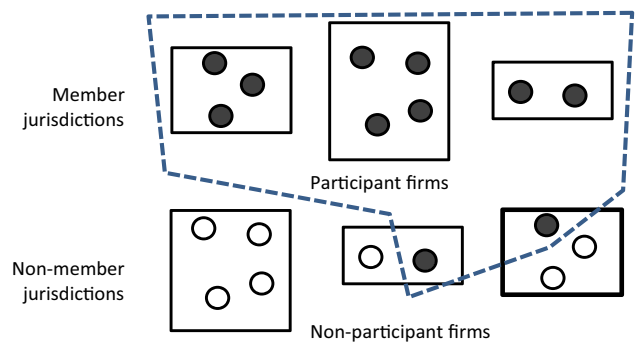

**Fig. 4** Schematic of membership and participation: some firms might participate though their jurisdiction is not a member

Consider the incentives of member jurisdictions who cheat. They might allow commercial agents to exceed production limits or they might allow non-participating agents to trade. Other participating agents would sanction the cheating agents and other member jurisdictions would sanction both the cheating jurisdiction and its cheating agents. These sanctions can include loss of committee membership. As more jurisdictions join, a cheating jurisdiction can be more easily sanctioned and with greater sanctions. The incentive for jurisdictions to participate appears to increase over time.

Consider a large country that does not participate as a whole but has participating jurisdictions within. Sprinz et al. [31] point out the hindrance from lack of U.S. leadership in a climate club. This hindrance would be loosened by participation of states and loosened further through participation by firms in non-participating states.

In addition to the incentives between jurisdictions and firms, and between firms, this design empowers individual consumers to harry non-participating firms directly by reduction in demand, or at least reputation (though reputational penalties are not always enough [32]), even if those firms are in nonmember jurisdictions. Citizen boycotts of non-participating firms would incentivize those firms to participate, picking them off one at a time, while inhibiting drop-out behavior. This strongly contrasts with the inability of consumers to influence fossil fuel emitters in the absence of a centralized mechanism.

These strong incentives to participate should go a long way toward eliminating free-riding behavior. While these behaviors need to be simulated to understand the effectiveness of the sanctions, the individual-to-business sanctions and business-to-business sanctions suggest that these incentives are a superset of and therefore at least as strong as those proposed by Nordhaus [15].

The double-sided nature of the auction might incentivize emitters to vertically integrate with sequestration activities. The incentives come from economies of scope where emitting firms could convert their emitting technologies to sequestering technologies and from economies of scale where emitting firms want to invest in any kind of sequestration to reduce permit price risk.

### 3.4 Objections and potential problems

What could go wrong?! Implementation of an ETS is affected by many factors, such as constitutional provisions, international treaties, the ability to create competition, the influence of emitters, public opinion, and the government's need to raise revenues [4]. All of these and others will apply to this proposal and I am not ready to address all of them. But here are a few.

#### 3.4.1 Uncertainty in the warming parameters $W_{p,u,t}$

The warming parameters $W_{p,u,t}$ in SMDAMAGE are uncertain and researchers have characterized the uncertainty [33, 34]. A production model of SMDAMAGE could use conservative values of $W_{p,u,t}$ to be, say, 99% sure that warming is within the constraint. This would be easy to implement in SMDAMAGE.

Most ETSs use global warming potential (GWP) to enable pricing of different GHGs as a function of the permit price of $CO_2$. For a kilogram of a selected GHG, GWP is the radiative forcing (standardized so the GWP of $CO_2$ is 1) over a time interval, e.g., 20 years or 100 years [34].

Using GWP in a price-based mechanism has several problems. First, the ETS must choose the time length for the GWP, i.e., implementers must choose whether to limit warming in the short-term or long-term, resulting in incorrectly capping long-term or short-term warming ([35] suggests the 20-year GWP is worse than the 100-year GWP). Second, except for a crude length in years, GWP ignores the timing of the heating over the selected period. Third, use of GWP raises costs through market inefficiency [36]. In an ETS with GWP factors, the permit price for a given GHG is its GWP times the $CO_2$ permit price. Yet Table 1 shows price ratios changing over time for a given pair of gasses. While researchers have attempted to find improved GWP factors [37], SMDAMAGE avoids the errors completely by using the warming function directly.

### 3.4.2 Fairness between countries, equity considerations

Determining whether a given country breaks even would require a numerical simulation of SMDAMAGE for each country. The large energy exporters are likely to have the greatest losses [38] unless they can pivot to sequestration.

Regarding equity, Stiglitz [14, 39] is concerned that carbon taxes will hurt the poor more than the rich. His papers therefore recommend both carbon taxes and regulation, such as requiring a particular sector to change to a greener technology, nonlinear electricity tariffs, and subsidies for public transportation.

In my view, global warming will hurt—is already injuring—the poor far more than carbon pricing will hurt them. In addition, resolving the problem of inequity adds large transaction costs (getting to agreement about equity) to the primary problem of reducing warming. If we instead move forward risking the appearance of hardheartedness, we have a better chance of solving the common problem and we are likely to reduce costs for the poor.

While the proposal here is more stringent than any previous, it does not exclude subsidies or regulations. Because it limits quantities, jurisdictions could subsidize affected commodities (at high risk of creating shortages), public transportation, etc.

More generally, SMDAMAGE has nothing explicit about the social cost of carbon. Pezzey [40] and Kaufman et al. [11] discard the social cost of carbon as impractical for policy use; this paper does the same. (For a sophisticated analysis, see Cai and Lontzek [41].) Implicitly, this proposal should lower the expected social cost of carbon and increase its certainty by increasing certainty about the warming pathway. Other researchers might figure out how to incorporate the social cost of carbon directly in the objective function of SMDAMAGE.

### 3.4.3 Legal impediments

Implementation of SMDAMAGE will require consideration of international treaties, constitutional provisions, and local laws. These seem surmountable, but others have more expertise in legal impediments. Gardoqui and Ramirez [42] discuss global trade rules in the context of a climate club. Weil [43] discusses these rules at greater length and pessimism.

### 3.4.4 Initial allowances

To introduce ETS allowances, governments can auction them or give them away [44]. Giving away permits lowers the recipient's incentive to abate emissions, entrenches them in past behavior and technology, and can give recipients windfall profits and a competitive advantage against new entrants; free permitting is subject to political manipulation [45]. Thurber and Wolak [46] found policymakers should minimize free allocations. The regulatory cap can result in price increases that raise participants' revenue [45].

Besides those disadvantages, free allocation of permits reduces the market manager's revenue, thus hindering purchase of sequestration contracts. This proposal rejects free allocation for a pure polluter-pays mechanism.

Given the excess inventory of atmospheric $CO_2$, initial allowances could be negative. Following the start of the market, the governing committee could require permits for all emissions after a past year, say 2010. The committee could use SMDAMAGE to calculate prices for those earlier emissions; those prices would likely be high. Further delay raises these costs. A credible threat to charge for past emissions could incentivize agents to reduce their emissions before the market begins.

### 3.4.5 Liquidity

According to Holt and Shobe [47], key measures of performance of the market include liquidity, price discovery, efficiency, and price volatility. SMDAMAGE should have excellent price discovery and economic efficiency. The next section discusses volatility. Liquidity is "a measure of how many units of the market asset…are available to come into the market as the price rises", as they put it. Holding limits intended to prevent market power reduce liquidity. But market limits are not desirable. A liquidity shortage could occur in early years if bids for emission cannot be offset by sufficient offers to sequester. The corresponding high prices would serve to kick-start behavior change.

### 3.4.6 Volatility

Policymakers and researchers are concerned about an ETS's ability to flex in response to economic shocks [48]. Doda [49] wrote, "A well-designed system can prevent prices from falling too low during a recession and so maintain the abatement incentive, or from overshooting in a boom and excessively constraining production by regulated firms precisely when they are at their most productive." Such flexibility seems unnecessary in general. Discarding market interventions such as price ceilings and floors decreases the management uncertainty about solving the problem. Markets should adjust on their own and governments are supposed to do little about it.

### 3.4.7 Competitiveness

ETSs can affect international competitiveness, but mitigating mechanisms can lower economic efficiency [44, 50–53]. They can suffer from agents with market power [54] and can inhibit project development [55]. Offsets can be ineffective [56], and regulators allow "self-protection" and sector-based targets inconsistent with price efficiency [28]. Some of these problems are likely associated with the different degrees of ETS implementation; apart from equity, the playing field is level if all countries face identical prices. Competitiveness problems could arise due to transfer prices for a multinational corporation with branches in countries with different local accounting rules.

### 3.4.8 Revenue sufficiency

Since neutral or positive revenue increases public support for cap and trade [57, 58], the potential lack of revenue neutrality, depending on the choice of $Cap_t$, could hinder its implementation. The governing committee should also accept donations from governments, foundations and individuals. Charitable institutions and individuals could donate funds to increase their prestige.

A simple mechanism to resolve the revenue sufficiency is to reduce the coefficients on sequestration. Reducing the coefficients on sequestration makes sense when the amount of sequestration is uncertain.

### 3.4.9 Moral hazard

Solutions to global warming are rife with adverse selection (hidden information affects the risk of transaction) and moral hazard (hidden behavior affects the risk of transaction).

Where to start? Inability to measure other countries' emissions [59], fantasizing of carbon capture technology [60], fantasizing of geo-engineering [61], fantasizing of fusion, fantasizing that technology cannot help at all [62], hoping everyone else will reduce emissions so we don't have to [63], hoping future generations will be richer and better able to solve the problem [64] as in arguing about the discount rate [65], producing a greenhouse chemical to get the credit for destroying it, threatening to clear the forest to get paid [66], clearing a forest to get a credit for planting, taking money to plant a forest that was going to be planted anyway [67, 68], taking money to plant the forest but not following through.

We cannot solve all the problems of moral hazard. The design here aims to avoid or solve some of them. At least this proposal should do no worse than existing ETSs.

Other people might be able to do better. The European ETS administrators have worked hard at this. Small sequestration contracts could be aggregated by third-parties with sufficient reputation to provide some guarantees. The member committee could choose to adjust SMDAMAGE coefficients $W_{p,u,t}$ for sequestration on the assumption that some contracts will fail. The governing committee could implement some kind of mechanism for clawing back money if the market does not meet its temperature targets, but the design for such a mechanism is an open problem. I have suggested a whistleblower mechanism. Drones can count trees now. The researchers in mechanism design should be able help. Petrakis and Xepapadeas [59] address moral hazard in measuring emissions. MacKenzie, Ohndorf and Palmer [69] have a hopeful proposal titled "Enforcement-proof contracts with moral hazard in precaution: Ensuring 'permanence' in carbon sequestration."

But we cannot fold our hands because we're afraid of the moral hazard. That moral hazard itself is the worst. The solution is for the market operator and member committee to improve and adjust over time.

## 4 Numerical results

The Appendix describes a numerical simulation of SMDAMAGE. While it is a cartoon, the simulation conveys the possible behavior of the market.

*The auction outcome will surprise.* In many scenarios and years, emissions of many GHGs were above 90% of their current emissions. However, the timing is critical. These same GHGs had low fractions of their maximum bids early in the auction horizon before the first constraining year. In these early years, the auction manager buys many contracts for sequestration by agriculture and forestry and accepts few bids for GHGs. The sequestration contracts take time to have enough effect. Once enough sequestration has been planted, global emissions could rise again later.

*Static quantities and prices are wrong.* Getting to zero (i.e., returning the global average temperature to some previous number, such as the average temperature in 1920) by a particular year will require a significant immediate ramp-up. Prices change over time. Indeed, the only year in which the warming constraint $Cap_t$ is binding is in the first year it is constrained; warming is above zero before that year, and below zero after that year. This result shows the fallacy of static limits in carbon tax and cap-and-trade mechanisms. Table 2 shows selected example prices from a hypothetical SMDAMAGE auction run in 2020, with the first constrained year $Y = 2050$. The table shows high prices in the early years, declining quickly in just a few decades. They are dynamic and not monotonic.

Kaufman et al. [11] show results with increasing prices for $CO_2$. Their results may be valid when $CO_2$ is curtailed slowly for the near future and their results depend on the type of gas (Table 2 shows $CH_4$ prices first increasing). The results here show some high prices rising slightly in the first decades but then prices fall significantly and the reason is plain. GHG prices should be high to curtail present emissions before sequestration has ramped up.

*Fixed global warming potentials (GWP) are wrong.* In an ETS with fixed GWP factors, the permit price for a given GHG is the GWP times the $CO_2$ permit price. Yet Table 1 shows changing price ratios for a given pair of gasses over time. Consider for example [$SF_6$ \$/t]/[$CO_2$ \$/t] from Table 1. In 2060, this ratio is 10,468; in 2105, the ratio is 4632. This shows the error of using "$CO_2$ equivalent" ratios in pricing; those ratios ignore the dynamic warming effects and target date for the global temperature.

The global temperature would rise and then fall. After the first year of constrained warming, as agriculture and forestry sequester $CO_2$, the global temperature could fall and even overshoot the $Cap_t$ limit, drawing down the initial excess $CO_2$. The global temperature is unlikely to reach some kind of stasis in the next hundred years, but SMDAMAGE provides the tool to manage the global temperature.

The numerical simulation shows trajectories in global temperature as a function of the first constrained year $Y$ (Fig. 5). Each curve corresponds to an auction run in 2020, with the first constrained year $Y$ varying from 2029 to 2100. The graph shows the year $Y$ where the series crosses the axis. Delaying the first constrained year results in a larger temperature increase, but lower cost to the auction manager.

These curves look optimistic, which might be due to my assumptions about the data. The model ignores the ocean inventory of $CO_2$ and other physical features. Nevertheless, consider the objective compared to other ETSs: this mechanism uses market forces to target warming directly.

*Auction revenue depends on the first constraint year.* The numerical example (see Appendix) characterizes how the auction manager's net revenue in 2020 would vary with the date of the first "net zero" year. Figure 6 demonstrates this for an auction held in 2020, where $Cap_t = \infty$ for $t < Y$ and $Cap_t = 0$ for $t \geq Y$, for $Y$ varying from 2029 to 2100. To be clear, the costs recorded in the graph are not the total cost to the global economy, but rather the net cost to the auction manager. Auction outflow is high in the early years but would quickly decline. This suggests a net zero goal of 2040 to 2050 if we implement SMDAMAGE soon. Continuing to increase the GHG burden raises the cost of solving the problem; every kilogram we emit now we must sequester later. This calculation shows the cost of the current excess inventory and the folly of continuing to increase that inventory. This problem is not a shortcoming of SMDAMAGE; rather, SMDAMAGE makes the problem explicit. It seems implausible that repairing environmental damage could be revenue positive.

Table 2 Auction prices (constant $US2020, no discounting) by period and type of contract, for an auction held in 2020, with first constraint period $Y = 2050$. The table omits intermediate periods for brevity

| Auction period | Loblolly pine 150-year contracts $/hectare | Black walnut 10-year contracts $/hectare | Ponderosa pine 10-year contracts $/hectare | Agric $CO_2$ $/t | $CO_2$ $/t | $CH_4$ $/t | $N_2O$ $/t | HFC134a $/t | HFC125 $/t | HFC143a $/t | CF4 $/t | $C_2F_6$ $/t | $SF_6$ $/t |
|---|---|---|---|---|---|---|---|---|---|---|---|---|---|
| 2020 | $3844 | $382 | $316 | $16 | − $16 | − $108 | − $2403 | − $6811 | − $17,012 | − $22,001 | − $29,516 | − $66,798 | − $80,343 |
| 2025 | $3594 | $387 | $319 | $16 | − $16 | − $131 | − $2410 | − $7546 | − $18,442 | − $22,001 | − $31,549 | − $66,798 | − $114,808 |
| 2030 | $3247 | $390 | $322 | $16 | − $16 | − $160 | − $2386 | − $9179 | − $19,595 | − $22,001 | − $29,516 | − $66,798 | − $103,742 |
| 2035 | $2790 | $389 | $321 | $16 | − $16 | − $195 | − $2330 | − $10,506 | − $21,025 | − $22,386 | − $21,089 | − $37,999 | − $91,410 |
| 2040 | $2384 | $363 | $300 | $16 | − $16 | − $231 | − $2239 | − $11,992 | − $21,612 | − $23,202 | − $29,516 | − $66,798 | − $114,808 |
| 2045 | $2125 | $209 | $173 | $14 | − $14 | − $244 | − $2012 | − $11,690 | − $19,308 | − $20,003 | − $21,089 | − $37,999 | − $80,343 |
| 2050 | $2053 | $156 | $128 | $6 | − $6 | − $104 | − $1326 | − $4887 | − $9215 | − $10,793 | − $12,662 | − $28,654 | − $45,531 |
| 2055 | $2043 | $157 | $130 | $7 | − $7 | − $23 | − $1050 | − $1477 | − $5194 | − $8410 | − $12,662 | − $18,927 | − $45,531 |
| 2060 | $2020 | $159 | $131 | $7 | − $7 | − $26 | − $1068 | − $1779 | − $6059 | − $9226 | − $10,629 | − $28,654 | − $68,930 |
| 2065 | $1972 | $161 | $133 | $7 | − $7 | − $29 | − $1079 | − $1779 | − $6647 | − $9226 | − $12,662 | − $24,054 | − $57,863 |
| 2070 | $1898 | $163 | $135 | $7 | − $7 | − $32 | − $1085 | − $1922 | − $6925 | − $9612 | − $10,629 | − $28,654 | − $45,531 |
| 2075 | $1808 | $165 | $137 | $7 | − $7 | − $37 | − $1100 | − $2526 | − $7513 | − $10,427 | − $12,662 | − $24,054 | − $57,863 |
| 2080 | $1701 | $167 | $138 | $7 | − $7 | − $43 | − $1110 | − $2971 | − $8378 | − $10,427 | − $12,662 | − $24,054 | − $57,863 |
| 2085 | $1579 | $169 | $140 | $7 | − $7 | − $51 | − $1113 | − $3416 | − $9522 | − $11,629 | − $12,662 | − $28,654 | − $68,930 |
| 2090 | $1457 | $173 | $143 | $7 | − $7 | − $61 | − $1106 | − $3701 | − $9800 | − $11,199 | − $8364 | − $18,927 | − $45,531 |
| 2095 | $1300 | $176 | $145 | $7 | − $7 | − $75 | − $1096 | − $4448 | − $10,665 | − $12,015 | − $12,662 | − $28,654 | − $45,531 |
| 2100 | $1099 | $179 | $148 | $7 | − $7 | − $93 | − $1082 | − $5338 | − $12,119 | − $12,015 | − $12,662 | − $28,654 | − $45,531 |
| 2105 | $818 | $180 | $148 | $7 | − $7 | − $115 | − $1047 | − $6227 | − $12,119 | − $12,015 | − $8364 | − $18,927 | − $34,465 |
| 2110 | $467 | $176 | $146 | $7 | − $7 | − $142 | − $984 | − $7562 | − $12,397 | − $12,015 | − $8364 | − $14,327 | − $34,465 |
| 2115 | $186 | $127 | $105 | $7 | − $7 | − $166 | − $882 | − $8451 | − $12,397 | − $11,199 | − $8364 | − $14,327 | − $34,465 |

**Fig. 5** Temperature trajectories over time by first constrained year *Y* (where the series first crosses the axis)

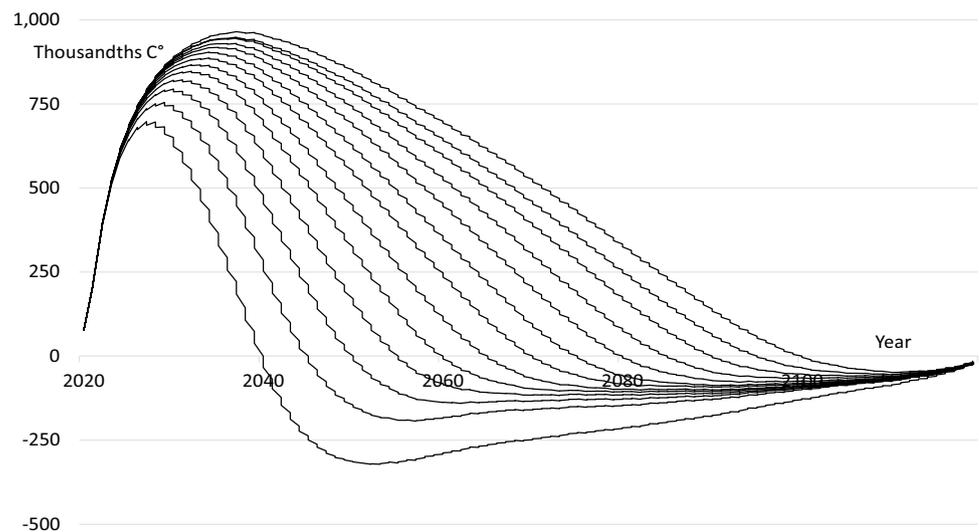

**Fig. 6** Auction revenue in 2020 by "net zero" year *Y*

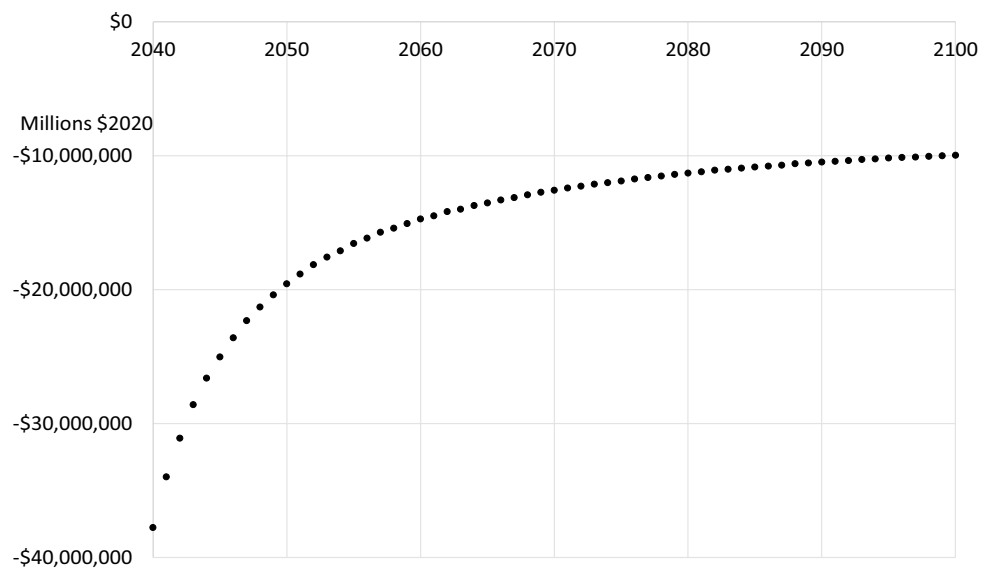

## 5 A faster path forward

### 5.1 Top-down or bottom-up?

This proposed market could be implemented in a top-down or bottom-up strategy. The top-down approach is a centralized effort to align the behavior of key decision makers [70]. But this faces an enormous transaction cost at jurisdictional level, so we do see few of these transactions.

Researchers have called for climate clubs of countries [15, 16, 70–72]. A club of ETSs could start SMDAMAGE. Such a club requires excludable benefits which the club can prevent non-members from enjoying, such as access to finance and border tariffs [72]. While not as difficult as a worldwide treaty, the transaction cost to develop a climate club is also large.

In contrast, transaction costs for actions by individuals, companies, and small jurisdictions are small, and we see many such transactions of this type. The bottom-up strategy has decentralized policy design and implementation [25, 70]. The bottom-up strategy requires institutions to encourage public and private entities to explore policy options and scale up successes. To be effective, though, the bottom-up strategy must evolve from decentralized scattered actions toward centralized coordinated actions.

So here is a new path for implementation of this proposal: willing businesses join to implement SMDAMAGE. Willing businesses are already reducing their emissions. Some jurisdictions will want to join to lower outlays for subsidies and regulations and to avoid implementing their own ETS. Monast [73] may call this proposal a "coordinated design strategy."

### 5.2 The key role of the firm on the permit side

The market rules and implementation path for SMDAMAGE suggest an inexorable path toward a global market.

A market maker takes the first step. This market maker could be a coalition of willing businesses or jurisdictions such as California. Most likely, businesses could start faster, so let's call the initial market maker GreenTechCo.

This market should save considerable money for firms with clean energy commitments. Bloomberg [30] recently reported that major Japanese firms may have to relocate factories due to lack of local clean energy. Purchasing offsets would likely be far cheaper. Executives would be assured of a first-class market mechanism. Further, a coalition lowers the fixed cost of operating the market mechanism by sharing that cost across firms.

GreenTechCo's coalition would establish the market institution as an independent nonprofit and hire the market manager. Let's call the market manager Greta. Greta, with oversight from the governing committee of members, establishes $Cap_t$ following Rule 1.

In step 2, the market maker lurches forward with demands that its suppliers, and the suppliers' suppliers, buy emissions permits from Greta as a condition of continued business with GreenTechCo. Let's call the supplier Emisor DeGas. GreenTechCo has incentive to help Emisor DeGas clean up its emissions. GreenTechCo's coalition invites, encourages, and pesters other companies to participate in the market. While GreenTechCo is willing to pay the cost of the emissions permits, GreenTechCo has incentive to push the responsibility for buying the permits upstream, so Emisor DeGas has incentive to reduce its emissions.

As steps 1 and 2 move to a walk, Greta begins to sell permits and begins to buy sequestration contracts. Standing up this work will be a major undertaking, but she has funding for it. More on this in the next section.

To add muscle to these legs, I propose the customer-labeled emissions permit (CLEP), an instrument I have not seen elsewhere (perhaps Nori LLC, Seattle, WA, is doing this). Think of a CLEP as a tag on the permit that Emisor DeGas buys from Greta. The tag shows the label "GreenTechCo," the period of production, the pollutant, and the quantity spewed by Emisor DeGas when it makes a batch of product for GreenTechCo. The contract registers the tag with Greta, who can verify its authenticity. Thus, Emisor DeGas has proof of permit specifically for GreenTechCo's materials, down to the shipment. The International Organization for Standardization could turn the CLEP into an ISO standard. With blockchain technology, every can of soup could have proof of permit. A customer could scan a QR code; the app would ping Greta's database. GreenTechCo can demand this proof for every shipment from Emisor DeGas. Thus, GreenTechCo can verify Emisor DeGas' adherence to Rule 2: participating commercial agents must get permits for all production. As a condition of a loan, a financial company can demand to see the CLEPs from an applicant borrower.

In step 3, consider Greta's point of view, and recall Rule 3 above: everyone has responsibilities to avoid trade with non-participating agents. Greta requires Emisor DeGas to get its own suppliers to buy permits. The jurisdiction of residence for the other suppliers does not matter. Thus, the market has a widening network of participating companies.

Greta can refine Rule 3 by defining "Bronze Participants" as firms who buy permits for all their emissions, but whose suppliers do not buy such permits. "Silver Participants" are firms who buy permits for all their emissions, and whose direct suppliers also buy permits, but not all suppliers in their supply chain participate in the SMDAMAGE market. "Gold Participants" have a supply chain in which all suppliers fully participate.

GreenTechCo should then have an easy time (compared to making a global treaty) convincing its headquarters city to begin demanding participation of its own suppliers, even if the city cannot require local firms to participate generally. A jurisdiction could itself be a bronze, silver, or gold participant, depending on what the jurisdiction requires of its business suppliers. A jurisdiction becomes a member when the jurisdiction passes a law to require all businesses within to participate in the SMDAMAGE market.

In step 4, recall Rule 4: the governing committee promotes individual concern worldwide to support participating companies and to boycott companies that do not. Consumers can insist on participation by their retail firms, and those retail firms can insist on participation by their suppliers. Rules 5 and 6 begin to impel the market forward, with Greta listing scofflaws and paying whistleblowers.

With step 5, willing jurisdictions join, following Rule 7: each member jurisdiction requires agents within its jurisdiction to participate.

In step 6, other jurisdictions find a lower political cost of membership, because many companies within are participating.

In step 7, the market moves to a sprint with Rules 9 and 10: member jurisdictions sanction non-member jurisdictions and non-participating firms.

These rules enable a divide-and-conquer strategy at the level of the individual firm because transaction costs are low.

- Jurisdictions can join one at a time without need for international treaty. They avoid standing up their own ETS administration, so small countries can join immediately. The jurisdiction has a low transaction cost to join, especially if firms within already participate.
- Firms can participate one at a time without their jurisdictions joining. Responsible CEOs have assurance that they are participating in the strongest known economic mechanism for reducing global warming. Their business partners have incentive to help them. Transaction costs of buying permits are lower with the central mechanism; a small company of goodwill can easily arrange for the permits it needs.
- Consumers can boycott one firm at a time whether or not their jurisdiction is a member. Early adopting firms win the race in marketing. Late adopting firms get mud on their face. Shareholders must demand action.

### 5.3 The sequestration side

Greta's business in selling permits will match her business in sequestration, but sequestration will take more work. Fortunately, she has funding for it and support from her member committee. Facing the moral hazards one at a time, she will have to help stand up supply chains of sequestration across the globe. Agricultural and forestry players might sign up quickly. Third parties may aggregate small contracts, selling them on behalf of small players. This will be messier than the permit side with higher costs for transactions and enforcement.

Biologists will have work in developing the SMDAMAGE coefficients $W_{p,u,t}$ (warming as a convolution of monthly carbon sequestration and the warming factor for $CO_2$) for the many types of sequestration, whether planting forest, planting grasses, planting mangroves, enhancing soil carbon, etc. Inspection firms will make hay in tracking sequestration.

## 6  Conclusion

In the short run, decision-makers are likely to want more research to understand a complete design of the market institution, the full role of the market manager, the auction schedule, the structure and nature of the bids, the bidding process, ways to simplify the bidding process for different types of economic agents, and automated bidding, and tools to mitigate adverse selection. Recognizing emissions as occurring at the same time as production could have unintended effects, where producers stock chemicals in advance of emission, anticipating higher later prices. The contract should at least specify a maximum time between production and emission. Other important issues for market design include the process of market clearing and the use of marginal cost pricing. The work is ripe for application of experimental economics, for gaming in the implementation process and in market operation, and for integration with partial equilibrium and other macroeconomic models.

Most likely, this numerical simulation is optimistic about temperature but pessimistic about the cost. A simulation less optimistic about temperature and less pessimistic about cost would shift the Fig. 6 graph rightwards and upwards. Estimated emissions as calculated by SMDAMAGE could be put back into a climate simulator for validation. On the other hand, bidding here ignores learning effects; bids were static over the full bidding horizon, implying business will always want to emit the same amount of GHGs. Discounting is ignored for simplicity of exposition, though the function is available in the code. Ignoring learning and discounting overestimates costs. Bids could be made more accurate by using marginal abatement curves by small region or city [74].

This smart market permit mechanism would extend to a range of other environmental applications. For example, jurisdictions could require manufacturers to pay for production of plastic while using the revenue to procure services for disposal and to clean up the oceans.

The road to implementation will be hard. Perhaps the greatest argument against this proposal is hopeless pessimism. Visionary leaders have solved hard problems in the past. Business people who believe in markets and individual choice, and who dislike government regulation and subsidies, would support this proposal. I call on the large tech firms to develop a consortium to implement SMDAMAGE.

It is easy to be cynical, but emissions are now widely traded and the problem has the world's attention. We all need certainty of the warming pathway, we need the lowest cost solution to this most expensive of problems, we need a robust mechanism that is implementable, and we need it soon. This paper has proposed a plausible path, for this world drunk on heat, as I wish it would go. We can easily imagine many ways people will break rules and cheat and stall and hinder and even write obstructive laws in bad faith. None of those is sufficient reason to stand still anymore.

**Acknowledgements**  I am indebted to the many researchers who have explained these complicated matters through their articles and who produced reams of data to share. I'm grateful to the open source software community for producing Python, the modeling language PuLP (J.S. Roy and S.A. Mitchell), the solver CBC (COIN-OR Foundation), and Hartin et al., [75] for the GHG simulator Hector. Thanks to Dr. Greg Shively and Prof. Kelly Klima for their help and advice. Thanks to Dr. Marygail Brauner, Prof. Craig Bond, and Prof. Keith Willett for their inspiration and suggestions on this document. Thanks to Mathijs Harmsen for his help with marginal abatement costs. Thanks to Dr. Marilyn K. Raffensperger for her help and support.

**Policy insights**  Existing emissions trading systems and carbon taxes do not constrain warming to any specific level, have poor incentives to join, and fail to produce useful future price information. This paper proposes a centralized market, cleared with optimization, for emission and sequestration contracts tied to temperature change. A coalition of firms would start the market institution and require their suppliers to participate; jurisdictions would join later. Unlike existing emissions trading systems, the market structure and rules incentivize firms to participate even if their jurisdiction does not, potentially resulting in a rush-to-join behavior.

**Authors' contributions**  I wrote the entire manuscript. This work is not a RAND output. I have done this on my own time. The author read and approved the final manuscript.

**Data availability**  Data and code for this study are available from the author on reasonable request.

**Competing interests**  The author declares no competing interests.

## Appendix: numerical simulation

This section describes a numerical simulation of SMDAMAGE. The results are speculative for a range of reasons. I had to speculate about global demand for emissions, global supply of sequestration, and a path of planned temperature. While much of this data is available publicly, my use of it might not have done justice to the authors' intent. Further, pulling together so much disparate data proved technical and complicated and I might have made mistakes in it.

### Activities simulated

The simulation omits all abatement activities with negative cost. This assumption is conservative in the sense that agents would abate a great amount at very low cost simply when faced with a price.

Even with its display of future prices, the numerical simulation is static: it does not take into account how agents will react to prices once posted. As agents react to the first round of prices, we would expect a change in the revealed abatement curves during bidding in the next period.

For each activity, the simulation requires:

- Bids values are from the literature on marginal abatement cost curves, as described below. Bid selection required assumptions about behaviors. For example, the available data on forestry provides a range of plantable total area, but little information about the rate at which foresters could plant. To convert from a static marginal cost curve to a set of bids over time, assume 20 years to plant the highest number of *H* hectares, so foresters could plant this area at a maximum rate of *H*/20 hectares per year. Thus, the change in behavior requires a ramp-up time. The optimization allowed 10-year, 20-year, and 150-year contracts for forest. An early simulation with only 150-year contracts proved costly and had considerable under-shoot in temperature. Simply adding more contract types gives the optimization more options and can lower cost. People will want to offer many contract types not imagined here, so the cost here is likely a wild over-estimate.
- Warming effects, ideally recorded as the degrees Celsius change in each period after the activity. These warming effects come from the climate simulation Hector [75]. To obtain the warming effects from Hector, I simulated a 50% increase in the base emission of each chemical, and then recorded the change in temperature for each future year in the simulation, to 281 years into the future. This simulation defined the warming effect $W_{p,u,t}$.

The simulation includes the following activities.

1. Agriculture, megatons $CO_2$ sequestered per period.
   I extracted bid data from [76, their Fig. 11.17]. Sorting and aggregating by price produced a three-point bid stack. US$2005 were inflated to $US2020 and fitted a curve: megatonnes $CO_2$/year mitigated $= 0.0587p^2 + 39.613p + 926.25$, where $p$ is $US2020 price. Then I produced 57 bids in $4 increments from $0 to $224/ton $CO_2$, to a maximum of 6854 megatons/year; assuming zero megatons would be sequestered at a price of $0. I further divided the bid quantity by the number of periods per year. The decision variables are in units of megatons, so an objective coefficient of $100 implies a bid of $100 million per megaton.
   Assume agriculture sequesters $CO_2$ only in the period of bidding. Warming effects are from Hector based on the agricultural $CO_2$ sequestered per period.
   Assume no initial stock of agriculture, although the initial burden of $CO_2$ calculated below implies an initial inventory of natural sequestration.
2. Forestry, million hectares planted per period, represented by three tree types labeled black walnut, loblolly pine, and Ponderosa pine, each with contract lengths of 10 years, 20 years, and 150 years.
   Sequestration data were extracted from [77, their Fig. 2], which shows the annual carbon uptake by year for loblolly pine, Ponderosa pine, and black walnut, in different regions of the United States. Summing over the 155 years in the graph gave total short tons $CO_2$/acre for each representative tree type, which is the basis for creating bids based on sequestration. I then converted from short tons/acre to metric tons/hectare. Finally, calculate $W_{p,u,t}$ for $p =$ loblolly pine, Ponderosa pine, and black walnut. The decision variables are in units of a thousand hectares per period. For a hectare of pine trees planted in period $u$, denote the schedule of tons $CO_2$ sequestered as $N_{u+0}$, $N_{u+1}$, $N_{u+2}$, $N_{u+3}$, …, $N_T$. Then a contract for planting a thousand hectares in period $u$ will result in warming of period $t$ by the following:

$$W_{\text{pine},u,t} = -(1000\,h) * \sum_{v=u}^{t} \left(N_v \text{mtons } CO_2\right)/(1000\,h)/\text{period} * \left[\left(W_{CO_2,v,t} 10^{-3} C°\right)/\left(\text{mtons } CO_2\right)\right]$$

To obtain bids for global forestry [77, their Fig. 6], shows estimates for $1997/ton costs for carbon sequestration per year. I fit a curve to match their graph to get $/ton as a function of (US) megatons $CO_2$/year. The highest value on their chart served as the global highest bid for forestry. I inflated from $1997 to $2020, converted from acres to hectares, and from short tons to metric tons. Finally, I multiplied the sequestration per hectare over the life of each contract type by the $/ton to get $2020/hectare.
   Extrapolating to the globe [78, Sect. 4.3.7.2], indicates the earth might have up to 500 million hectares available for forestry. Assume the 500 million hectares had a maximum planting rate of 25 million hectares per year, equally divided among the three types of representative trees. Bids then ranged linearly in steps of 25,000 hectares per tree type per year, up to a maximum of 8,000,500 hectares/year, and from a low of $1724/hectare of loblolly pine to a maximum of $82,028/hectare of ponderosa pine.
   This estimate is speculative, in part because it builds up the bids based on a priori estimates of the value of carbon sequestration, not the cost of planting and operating a forest, and because the simulation uses only three representative tree types common to the U.S.
   Further, the simulation does not model the total capacity of forestry. Bid quantities are small enough that ramp-up in the model takes at least 20 years, but does not limit the total forestry area. Fortunately, the solution never uses all the forestry area.
   The simulation assumes zero inventory of forestry, although the calculation for the initial burden of $CO_2$ implies an initial inventory of natural sequestration.
3. $CO_2$, megatons (GtC in Hector) emitted per period. Warming effects from Hector (as "ffi").

    Bid data come from [79], who give marginal abatement cost coefficients in €2005 by country for energy-intensive and non-energy-intensive sectors. I compiled the marginal abatement cost curves for both sectors and all countries into a single bid stack in 10 megaton increments, with an upper bound of €1000, and then converted prices to $US2020 (at €1/$1.1, and inflating from 2005 to 2020).

    As the initial burden, I calculated the megatons $CO_2$ in the atmosphere at 410 ppm for 2020, and subtracted the calculated megatons $CO_2$ in the atmosphere at 303 ppm for 1920, resulting in a 2020 burden of 810,958 megatons of $CO_2$. The governing committee may prefer a baseline year other than 1920.

4. CH4 and N20, megatons emitted per period. Warming effects from Hector. For both of these, bid data came from [80], and developed bids from a curve fitted to their year 2100 emissions curve.

    For these chemicals and the remaining, an initial burden was based on the current atmospheric concentrations, from [81–84].

5. C2F6, kilotons emitted per period. Warming effects from Hector.
6. CF4, kilotons emitted per period. Warming effects from Hector.
7. HFC125, kilotons emitted per period. Warming effects from Hector.
8. HFC134a, kilotons emitted per period. Warming effects from Hector.
9. HFC143a, kilotons emitted per period. Warming effects from Hector.
10. $SF_6$, kilotons emitted per period. Warming effects from Hector.

## Software

I wrote the model in Python with the open source modeling language PuLP and the COIN-OR CBC solver. The model typically solved in 3 or 4 min.

## Running the simulation

Following assembly of the data and debugging of the model, the model was set to have its first auction in 2020, with bidding through $T_B = 2170$, and the last constrained period as $T = 2301$, with two bid periods per year.

    The simulation constrains the change in temperature to zero starting in $Y = 2020$, so $Cap_{2020} = 0.0$. This model was, not surprisingly, infeasible. A second run minimized temperature. The solution accepted all bids for sequestration and none for any emissions. Temperature rose a half degree C through 2024, but lowered below zero by mid-year 2028. This implausibly optimistic scenario set the limits on the mathematical solution space.

    I then ran 81 simulated auctions. In each auction, the market opened in 2020 with the first warming-constrained year ($Cap_Y = 0.0$) in year $Y$, for $Y = 2020$ to 2100. The model was infeasible for $Y = 2020$ to 2028, but feasible for $Y = 2029$ to 2100.

## References

1. World Bank 2019. Carbon pricing dashboard. https://carbonpricingdashboard.worldbank.org/map_data. Accessed 26 May 2019.
2. Wilcoxen Peter J, McKibbin WJ. Climate change after Kyoto: a blueprint for a realistic approach. The Brookings Institution. 2002. https://www.brookings.edu/articles/climate-change-after-kyoto-a-blueprint-for-a-realistic-approach. Accessed 6 Sept 2020.
3. Haya B. Policy brief: The California Air Resources Board's U.S. Forest offset protocol underestimates leakage. Univ. of California, Berkeley, 7 May 2019. https://gspp.berkeley.edu/assets/uploads/research/pdf/Policy_Brief-US_Forest_Projects-Leakage-Haya_2.pdf.
4. Haites E. Carbon taxes and greenhouse emissions trading systems: what have we learned? Clim Policy. 2018;18:955–66.
5. Farquharson DeVynne, Jaramillo P, Schivley G, Klima K, Carlson D, Samaras C. Beyond global warming potential: a comparative application of climate impact metrics for the life cycle assessment of coal and natural gas based electricity. J Ind Ecol. 2016;21:4.
6. Edwards MR, Trancik JE. Climate impacts of energy technologies depend on emissions timing. Nat Clim Change. 2014;4:347–52.
7. World Bank. State and trends of carbon pricing 2019. Washington DC: World Bank Group; 2019. p. 2019.
8. Stavins RN. The future of U.S. carbon pricing policy. M-RCBG Faculty working paper series, Mossavar-Rahmani Center for Business & Government, Harvard Kennedy School. 2019.
9. Shultz GP, Halstead T. The pricing advantage. Climate Leadership Council. 2020.
10. Weisbach DA, Metcalf GE. The design of a carbon tax. Harv Environ Law Rev. 2009;33:499.
11. Kaufman N, Barron AR, Krawczyk W, Marsters P, McJeon H. A near-term to net zero alternative to the social cost of carbon for setting carbon prices. Nat Clim Change. 2020. https://doi.org/10.1038/s41558-020-0880-3.
12. Cramton P, MacKay DJC, Ockenfels A, Stoft S. Global carbon pricing: the path to climate cooperation. Cambridge: MIT Press; 2017.
13. Tvinnereim E, Mehling M. Carbon pricing and deep decarbonisation. Energy Policy. 2018;121:185–9.
14. Stiglitz JE. Addressing climate change through price and non-price interventions. NBER Working Paper 25939, National Bureau of Economic Research, Cambridge, MA 02138, June 2019.

15. Nordhaus WD, Clubs C. Overcoming free-riding in international climate policy. Am Econ Rev. 2015;105(4):1339–70.
16. Nordhaus W. The climate club, how to fix a failing global effort. foreign affairs. May/June 2020, https://www.foreignaffairs.com/articles/united-states/2020-04-10/climate-club. Accessed 7 Sept 2020.
17. Weitzman ML. Prices vs. quantities. Rev Econ Stud. 1974;41(4):477–91.
18. McCabe KA, Rassenti SJ, Smith VL. Smart computer-assisted markets. Science. 1991;254(5031):534–8.
19. Abrell J, Rauch S. Combining price and quantity controls under partitioned environmental regulation. Report 301, MIT Joint Program on the Science and Policy of Global Change, July 2016.
20. Marques G, de Godoy S, Macchione Saes MS. Cap-and-trade and project-based framework: how do carbon markets work for greenhouse emission reduction? Ambient Sociedade. 2015. https://doi.org/10.1590/1809-4422ASOC795V1812015en.
21. Heindl P. The impact of administrative transaction costs in the EU emissions trading system. Clim Policy. 2017;17(3):314–29.
22. Dantzig GB. Linear programming and extensions. Princeton: Princeton University Press; 1998.
23. Shine KP, Fuglestvedt JS, Hailemariam K, Stuber N. Alternatives to the global warming potential for comparing climate impacts of emissions of greenhouse gases. Clim Change. 2005;68(3):281–302.
24. Vogt-Schilb A, Hallegatte S. Marginal abatement cost curves and the optimal timing of mitigation measures. Energy Policy. 2014;66:645–53. https://doi.org/10.1016/j.enpol.2013.11.045hal-00916328.
25. Vasconcelos VV, Hannam PM, Levin SA, Pacheco JM. Coalition-structured governance improves cooperation to provide public goods. arXiv, General Economics, 1910.11337. 2019.
26. Coria J, Jaraite J. Carbon pricing: transaction costs of emissions: trading vs. carbon taxes. Working Papers in Economics, No. 609, University of Gothenburg. 2015.
27. Crals E, Vereeck L. Taxes, tradable rights and transaction costs. Eur J Law Econ. 2005;20(2):199–223.
28. Gledhill R, Grant J, Low LP. Review of carbon markets. Briefing paper, PricewaterhouseCoopers LLP, UK and the Climate Group. 2008
29. Michaelowa A, Stronzik M, Eckermann F, Hunt A. Transaction costs of the Kyoto mechanisms. Clim Policy. 2003;3(3):261–78.
30. Bullard N. Japan Had better find some clean energy or risk losing its biggest businesses. Bloomberg. 2020. https://www.bloomberg.com/news/articles/2020-12-03/japan-had-better-find-some-clean-energy-or-risk-losing-its-biggest-businesses?srnd=premium. Accessed 3 Dec 2020.
31. Sprinz DF, Sælen H, Underdal A, Hovi J. The effectiveness of climate clubs under Donald Trump. Clim Policy. 2018;18:7.
32. Karpoff JM, Lott JR, Wehrly EW. The reputational penalties for environmental violations: empirical evidence. J Law Econ. 2005;48(2):653–75.
33. Lee LA, Reddington CL, Carslaw KS. On the relationship between aerosol model uncertainty and radiative forcing uncertainty. PNAS. 2016;113(21):5820–7.
34. Forster P, Ramaswamy V, Artaxo P, Berntsen T, Betts R, Fahey DW, Haywood J, Lean J, Lowe DC, Myhre G, Nganga J, Prinn R, Raga G, Schulz M, Van Dorland R. 2007: Changes in atmospheric constituents and in radiative forcing. In: Solomon S, Qin D, Manning M, Chen Z, Marquis M, Averyt KB, Tignor M, Miller HL, editors. Climate change 2007: the physical science basis. Contribution of Working Group I to the fourth assessment report of the intergovernmental panel on climate change. Cambridge: Cambridge University Press.
35. Climate Analytics. Why using 20-year Global Warming Potentials (GWPs) for emission targets are a very bad idea for climate policy. 2017. www.climateanalytics.org. https://climateanalytics.org/media/20-year_gwps_bad_idea_for_climate_policy_16112017.pdf.
36. Johansson DJA, Martin Persson U, Azar C. The cost of using global warming potentials: analysing the trade off between $CO_2$, $CH_4$ and $N_2O$. Clim Change. 2006;77:291–309.
37. Cain M, Lynch J, Allen MR, Fuglestvedt JS, Frame DJ, Macey AH. Improved calculation of warming-equivalent emissions for short-lived climate pollutants. Clim Atmos Sci. 2019;2:29.
38. Jacoby HD, Babiker MH, Paltsev S, Reilly JM. Sharing the burden of GHG reductions. Report No. 167, MIT Joint Program on the Science and Policy of Global Change. 2008.
39. Stiglitz J, Stern N, and the members of the High-Level Commission on Carbon Prices. Report of the high-level commission on carbon prices. Carbon Pricing Leadership Coalition, International Bank for Reconstruction and Development and International Development Association/The World Bank, May 29, 2017.
40. Pezzey JCV. Why the social cost of carbon will always be disputed. WIREs Clim Change. 2018;10(1):e558.
41. Cai Y, Lontzek TS. The social cost of carbon with economic and climate risks. J Polit Econ. 2019;127:6.
42. Gardoqui BL, Ramírez I. Identifying a WTO exception to incorporate climate clubs. BioResource. 9(7).
43. Weil G. Incentive compatible climate change mitigation: moving beyond the pledge and review. Model, 42 William & Mary Environmental. Law and Policy Review. 923; 2018. https://scholarship.law.wm.edu/wmelpr/vol42/iss3/6.
44. Goulder LH, Schein A. Carbon taxes vs cap and trade: a critical review. Clim Change Econ. 2013;4:1350010.
45. Huber BR. How did RGGI do it? Political economy and emissions auctions. Ecol Law Q. 2013;50(1):59–106.
46. Thurber MC, Wolak FA. Carbon in the classroom: lessons from a simulation of California's electricity market under a stringent cap-and-trade system. Electr J. 2013;26(7):8–21.
47. Holt C, Shobe W. Investigation of the effects of emission market design on the market-based compliance mechanism of the California cap on greenhouse gas emissions, No 2013-01, Reports, Center for Economic and Policy Studies. 2013. https://EconPapers.repec.org/RePEc:vac:report:rpt13-01.
48. Kollenberg S, Taschini L. Emissions trading systems with cap adjustments. J Environ Econ Manage. 2016;80:20–36.
49. Doda B. How to price carbon in good times… and bad. WIREs Clim Change. 2016;7(1):135–44.
50. Schmalensee R, Stavins RN. Lessons learned from three decades of experience with cap and trade. Rev Environ Econ Policy. 2017;11:59–79.
51. Schmalensee R, Stavins RN. The design of environmental markets: what have we learned from experience with cap and trade? Oxf Rev Econ Policy. 2017;33:572–88.
52. Fell H, Maniloff P. Leakage in regional environmental policy: the case of the regional greenhouse gas initiative? J Environ Econ Manag. 2018;87:10–23.
53. Fowlie M. Updating the allocation of greenhouse gas emissions permits in a federal cap-and-trade program. In: Fullerton D, Wolfram C, editors. The design and implementation of the U.S. climate policy. Chicago: University of Chicago Press; 2012.
54. Godby R. Market power in laboratory emission permit markets. Environ Resour Econ. 2002;23(3):279–318.

55. Larson DF, Breustedt G. Will markets direct investments under the kyoto protocol? Lessons from the activities implemented jointly pilots. Environ Resour Econ. 2009;43(3):433–56.
56. Wara MW, Victor DG. A realistic policy on international carbon offsets. Program on Energy and Sustainable Development, Working Paper No. 74. 2008.
57. Mills S, Rabe BG, Borick CP. Cap-and-trade support linked to revenue use. Issues Energy Environ Policy. 2015;23.
58. Klenert D, Mattauch L, Combet E, Edenhofer O, Hepburn C, Rafaty R, Stern N. Making carbon pricing work for citizens. Nat Clim Change. 2018;8:669–77.
59. Petrakis E, Xepapadeas A. Environmental consciousness and moral hazard in international agreements to protect the environment. J Public Econ. 1996;50(1):95–110.
60. Hamilton C. Moral Haze Clouds Geoengineering. EuTRACE Journal, Essay No. 1. 2013. https://www.iass-potsdam.de/sites/default/files/2018-04/hamilton_final_15.04.2013.pdf.
61. Fairbrother M. Geoengineering, moral hazard, and trust in climate science: evidence from a survey experiment in Britain. Clim Change. 2016;139:477–89.
62. Wagner G, Zizzamia D. Green moral hazards. NYU Wagner School of Public Service Research Paper Series. 2019.
63. Anesi V. Moral hazard and free riding in collective action. Soc Choice Welf. 2008;32:197–218.
64. Sachs JD. Climate change and intergenerational well-being. In: Oxford handbook of the macroeconomics of global warming. Oxford: Oxford University Press; 2015.
65. Nordhaus WD. The 'Stern Review' on the economics of climate change. NBER Working Paper Series, Working Paper 12741, National Bureau of Economic Research. 2006.
66. Fearnside PM. Brazil's Amazon forest in mitigating global warming: unresolved controversies. Clim Policy. 2012;12(1):70–81.
67. Burke PJ. Undermined by adverse selection: Australia's direct action abatement subsidies. Econ Pap J Appl Econ Policy. 2016;35:3.
68. Bushnell JB. Adverse selection and emissions offsets. Economics working papers (2002–2016). Iowa State University. 2011.
69. MacKenzie IA, Ohndorf M, Palmer C. Enforcement-proof contracts with moral hazard in precaution: ensuring 'permanence' in carbon sequestration. Oxf Econ Pap. 2011;64(2):350–74.
70. Sabel CF, Victor DG. Governing global problems under uncertainty: making bottom-up climate policy work. Clim Change. 2017;144(1):15–27.
71. Keohane N, Petsonk A, Hanafi A. Towards a club of carbon markets. Clim Change. 2017;144:81–95.
72. Paroussos L, Mandel A, Fragkiadakis K, Fragkos P, Hinkel J, Vrontisi Z. Climate clubs and the macro-economic benefits of international cooperation on climate policy. Nat Clim Change. 2019;9:542–6.
73. Monast J. From top-down to bottom-up climate policy: new challenges in carbon market design. San Diego J Clim Energy Law. 2016–17;8:175.
74. Ibrahim N, Kennedy C. A methodology for constructing marginal abatement cost curves for climate action in cities. Energies. 2016;9:227.
75. Hartin CA, Patel P, Schwarber A, Link RP, Bond-Lamberty BP. A simple object-oriented and open-source model for scientific and policy analyses of the global climate system—Hector v1.0. Geosci Model Dev. 2015;8:939–55.
76. Smith P, Bustamante M, Ahammad H, Clark H, Dong H, Elsiddig EA, Haberl H, Harper R, House J, Jafari M, Masera O, Mbow C, Ravindranath NH, Rice CW, Robledo Abad C, Romanovskaya A, Sperling F, Tubiello F. Agriculture, forestry and other land use (AFOLU). In: Edenhofer O, Pichs-Madruga R, Sokona Y, Farahani E, Kadner S, Seyboth K, Adler A, Baum I, Brunner S, Eickemeier P, Kriemann B, Savolainen J, Schlömer S, von Stechow C, Zwickel T, Minx JC, editors. Climate change 2014: mitigation of climate change. Contribution of Working Group III to the fifth assessment report of the intergovernmental panel on climate change. Cambridge: Cambridge University Press; 2014.
77. Stavins RN, Richards KR. The cost of US-based carbon sequestration. Pew center on global climate change. Cambridge: Harvard University; 2005.
78. de Coninck, H., A. Revi, M. Babiker, P. Bertoldi, M. Buckeridge, A. Cartwright, W. Dong, J. Ford, S. Fuss, J.-C. Hourcade, D. Ley, R. Mechler, P. Newman, A. Revokatova, S. Schultz, L. Steg, and T. Sugiyama, 2018: Strengthening and Implementing the Global Response. Global Warming of 1.5°C. An IPCC Special Report on the impacts of global warming of 1.5°C above pre-industrial levels and related global greenhouse gas emission pathways, in the context of strengthening the global response to the threat of climate change, sustainable development, and efforts to eradicate poverty [Masson- Delmotte, V., P. Zhai, H.-O. Pörtner, D. Roberts, J. Skea, P.R. Shukla, A. Pirani, W. Moufouma-Okia, C. Péan, R. Pidcock, S. Connors, J.B.R. Matthews, Y. Chen, X. Zhou, M.I. Gomis, E. Lonnoy, T. Maycock, M. Tignor, and T. Waterfield (eds.)]. In Press.
79. Anger N, Sathaye J. Reducing deforestation and trading emissions: economic implications for the post-kyoto carbon market. Discussion Paper No. 08-016, Zentrum für Europäische Wirtschaftsforschung GmbH. ftp://ftp.zew.de/pub/zew-docs/dp/dp08016.pdf. Accessed 30 Mar 2008.
80. Harmsen JHM, van Vuuren DP, Nayak DR, Hof AF, Höglund-Isaksson L, Lucas PL, Nielsen JB, Smith P, Stehfest E. Long-term marginal abatement cost curves of non-CO2 greenhouse gases. Environ Sci Policy. 2019;99:136–49. https://doi.org/10.1016/j.envsci.2019.05.013.
81. Ehhalt D, Prather M, Dentener F, Derwent R, Dlugokencky EJ, Holland E, Isaksen I, Katima J, Kirchhoff V, Matson P, Midgley P, Wang M, Berntsen T, Bey I, Brasseur G, Buja L, Collins WJ, Daniel JS, DeMore WB, Derek N, Dickerson R, Etheridge D, Feichter J, Fraser P, Friedl R, Fuglestvedt J, Gauss M, Grenfell L, Grubler A, Harris N, Hauglustaine D, Horowitz L, Jackman C, Jacob D, Jaegle L, Jain AK, Kanakidou M, Karlsdottir S, Ko M, Kurylo M, Lawrence M, Logan JA, Manning M, Mauzerall D, McConnell J, Mickley LJ, Montzka S, Muller JF, Olivier J, Pickering K, Pitari G, Roelofs G-J, Rogers H, Rognerud B, Smith SJ, Solomon S, Staehelin J, Steele P, Stevenson DS, Sundet J, Thompson A, van Weele M, von Kuhlmann R, Wang Y, Weisenstein DK, Wigley TM, Wild O, Wuebbles DJ, Yantosca R, Joos F, McFarland M. Atmospheric chemistry and greenhouse gases. Chapter 4 of the IPCC third assessment report climate change 2001: the scientific basis, 1 Oct 2001.
82. Ravishankara AR, Daniel JS, Portmann RW. Nitrous oxide (N2O): the dominant ozone-depleting substance emitted in the 21st century. Science. 2009;326(5949):123–5.
83. Tonkovich G. Air quality, energy, and greenhouse, gas emissions impact analysis, talbert extraction well decommissioning project. Vista Environmental, Project No. 19023, 6 May 2019.

84. Prinn RG, Weiss RF, Fraser PJ, Simmonds PG, Cunnold DM, Alyea FN, O'Doherty S, Salameh P, Miller BR, Huang J, Wang RHJ, Hartley DE, Harth C, Steele LP, Sturrock G, Midgley PM, McCulloch A. A history of chemically and radiatively important gases in air deduced from ALE/GAGE/AGAGE. J Geophys Res. 2000;105(17):17751–92.

**Publisher's Note**  Springer Nature remains neutral with regard to jurisdictional claims in published maps and institutional affiliations.